\documentclass[10pt]{article}
\usepackage{graphicx}
\usepackage{amsmath}
\usepackage{amssymb}
\usepackage{caption2}
\begin{document}
\bibliographystyle{prsty}
\begin{center}
{\large {\bf \sc{ Scalar tetraquark state candidates: $X(3915)$, $X(4500)$ and $X(4700)$ }}} \\[2mm]
Zhi-Gang Wang \footnote{E-mail: zgwang@aliyun.com.  }       \\
 Department of Physics, North China Electric Power University, Baoding 071003, P. R. China
\end{center}

\begin{abstract}
In this article, we tentatively assign the  $X(3915)$ and $X(4500)$ to be the ground state and the first radial excited state of the axialvector-diquark-axialvector-antidiquark  type scalar $cs\bar{c}\bar{s}$ tetraquark states, respectively, assign the $X(4700)$ to be the ground state vector-diquark-vector-antidiquark type scalar $cs\bar{c}\bar{s}$ tetraquark state, and study their  masses and pole residues with the QCD sum rules in details  by calculating the contributions of the vacuum condensates up to dimension 10. The numerical results support  assigning   the  $X(3915)$ and $X(4500)$ to be the ground state and the first radial excited state of the axialvector-diquark-axialvector-antidiquark type scalar $cs\bar{c}\bar{s}$ tetraquark states, respectively, and assigning  the $X(4700)$ to be the ground state vector-diquark-vector-antidiquark type scalar $cs\bar{c}\bar{s}$ tetraquark state.
\end{abstract}

 PACS number: 12.39.Mk, 12.38.Lg

Key words: Tetraquark  state, QCD sum rules

\section{Introduction}

In 2009, the $X(4140)$ was first observed by  the CDF collaboration   in the $J/\psi\phi$ mass spectrum in the $B^+\rightarrow J/\psi\,\phi K^+$ decays with a statistical significance in excess of $3.8 \sigma$  \cite{CDF0903}.
 In 2011, the CDF collaboration confirmed the
$Y(4140)$ in the $B^\pm\rightarrow J/\psi\,\phi K^\pm$ decays  with
 a statistical significance greater  than $5\sigma$, and  observed an evidence for the new resonance  $X(4274)$ with an approximate statistical significance of $3.1\sigma$
\cite{CDF1101}.
In 2013, the CMS collaboration   confirmed the $X(4140)$ in the $J/\psi\phi$  mass spectrum in the $B^\pm \to J/\psi \phi K^\pm$ decays,  and fitted the structure to an S-wave relativistic Breit-Wigner line-shape above a three-body phase-space nonresonant
component with a statistical significance exceeding $5 \sigma$ \cite{CMS1309}.
In the same year,  the D0 collaboration confirmed  the  $X(4140)$  in the  $B^+ \to J/\psi \phi K^+$ decays with a  statistical significance of $3.1\sigma$  \cite{D0-1309}.

Recently, the LHCb collaboration performed the first full amplitude analysis of the $B^+\to J/\psi \phi K^+
$   decays with $J/\psi\to\mu^+\mu^-$, $\phi\to K^+K^-
$  with a data sample of 3 $\rm{fb}^{-1}$ of  $pp$ collision data collected at $ \sqrt{s}=7$ and  $8\,\rm{TeV}$ with the LHCb detector,
confirmed the $X(4140)$ and $X(4274)$ in the $J/\psi \phi$ mass spectrum with statistical significances of  $8.4\sigma$ and $6.0\sigma$, respectively,  and determined the    spin-parity to be $J^{P} =1^{+}$ with statistical significances of $5.7\sigma$ and $5.8\sigma$, respectively \cite{LHCb2016}. Moreover, the LHCb collaboration observed  the new particles  $X(4500)$ and $X(4700)$ in the $J/\psi \phi$ mass spectrum with statistical significances of $6.1\sigma$ and $5.6\sigma$, respectively,  and determined the    spin-parity to be $J^{P} =0^{+}$ with statistical significances of $4.0\sigma$ and $4.5\sigma$, respectively \cite{LHCb2016}. The measured  masses and widths are
\begin{flalign}
 & X(4140) : M = 4146.5 \pm 4.5 ^{+4.6}_{-2.8} \mbox{ MeV}
\, , \, \Gamma = 83 \pm 21 ^{+21}_{-14} \mbox{ MeV} \, , \nonumber \\
 & X(4274) : M = 4273.3 \pm 8.3 ^{+17.2}_{-3.6} \mbox{ MeV}
\, , \, \Gamma = 56 \pm 11 ^{+8}_{-11} \mbox{ MeV} \, , \nonumber \\
 & X(4500) : M = 4506 \pm 11 ^{+12}_{-15} \mbox{ MeV} \, ,
\, \Gamma = 92 \pm 21 ^{+21}_{-20} \mbox{ MeV} \, , \nonumber \\
 & X(4700) : M = 4704 \pm 10 ^{+14}_{-24} \mbox{ MeV} \, ,
\, \Gamma = 120 \pm 31 ^{+42}_{-33} \mbox{ MeV} \, .
\end{flalign}
 The $X(4140)$, $X(4274)$, $X(4500)$ and $X(4700)$ are all observed in the $J/\psi\phi$ mass spectrum, if they are tetraquark states,  their quark constituents must be $cs\bar{c}\bar{s}$. The S-wave   $J/\psi\phi$ systems  have the quantum numbers $J^{PC}=0^{++}$, $1^{++}$, $2^{++}$,  the P-wave $ J/\psi\phi$ systems have the quantum numbers $0^{-+}$, $1^{-+}$, $2^{-+}$, $3^{-+}$ \cite{Review-Jpsiphi}. We can construct the interpolating  currents with $J^{PC}=1^{++}$ and $0^{++}$ to study the $X(4140)$, $X(4274)$ and $X(4500)$, $X(4700)$, respectively.

In Ref.\cite{WangHuangTao-3900},  we  study the masses and pole residues of the $J^{PC}=1^{+\pm}$ hidden charmed tetraquark states with the QCD sum rules. The theoretical predictions support assigning    the $X(3872)$ and $Z_c(3900)$  to be the $1^{++}$
and $1^{+-}$ diquark-antidiquark type tetraquark states, respectively. If we take the $X(4140)$ as the hidden strange cousin of the $X(3872)$, then $M_{X(4140)}-M_{X(3872)}=275\,\rm{MeV}$, the $SU(3)$ breaking effect is about $m_s-m_q=135\,\rm{MeV}$, which is consistent with our naive expectation.  However, detailed analysis based on the QCD sum rules indicates that it is unreasonable  to assign  the $X(4140)$ to be the diquark-antidiquark type   $cs\bar{c}\bar{s}$ tetraquark state with $J^{PC}=1^{++}$ \cite{WangX4140-1607}.

The   charged resonances  $Z^\pm_c(3900)$ and $Z^\pm(4430)$ have analogous decays \cite{PDG},
$Z_c(3900)^\pm \to J/\psi\pi^\pm$, $Z(4430)^\pm \to \psi^\prime\pi^\pm$.
The mass gaps are $M_{Z(4430)}-M_{Z_c(3900)}=576\,\rm{MeV}$ and $M_{\psi^\prime}-M_{J/\psi}=589\,\rm{MeV}$, so it is natural to assign
 the  $Z(4430)$ to be the first radial
excitation of the $Z_c(3900)$ \cite{Z4430-1405,Nielsen-1401,Wang4430}.
In Ref.\cite{Wang4430},  we  study the $Z_c(3900)$ and $Z(4430)$ with the QCD sum rules in details,  the theoretical predictions  support assigning  the $Z_c(3900)$ and $Z(4430)$
 to be the ground state and the first radial excited state of the $1^{+-}$  tetraquark states, respectively. Now we can draw the conclusion tentatively that the energy gap between the ground state and the first radial excited state of the tetraquark states is about $0.6\,\rm{GeV}$.

In 2004, the  Belle collaboration  observed  the $X(3915)$  in the $\omega J/\psi$  mass spectrum in the
 exclusive $B \to K \omega J/\psi$ decays \cite{Belle2004}. In 2007, the BaBar collaboration confirmed the $X(3915)$ in the $\omega J/\psi$  mass spectrum in the  exclusive $B \to K \omega J/\psi$ decays \cite{BaBar2007}. In 2010, the   Belle collaboration confirmed the $X(3915)$    in the two-photon process $\gamma\gamma\to \omega J/\psi$ \cite{Belle2010}. Now the $X(3915)$ is listed in the Review of Particle Physics as the $\chi_{c0}({\rm 2P})$
state with the quantum numbers $J^{PC}=0^{++}$ \cite{PDG}.
 In Ref.\cite{Lebed-3915},  Lebed and Polosa  propose that the  $X(3915)$ is the lightest
 $cs\bar{c} \bar{s}$ scalar tetraquark state based on  lacking of the observed $D\bar D$ and $D^*\bar{D}^*$
decay modes, and attribute  the single known decay mode $J/\psi \omega$ to the $\omega-\phi$ mixing effect.

If the  mass gap between the ground state and the first radial excited state of the tetraquark states is about $0.6\,\rm{GeV}$, just like in the case of the $Z_c(3900)$ and $Z(4430)$, the $X(4500)$ can be assigned to be  the first  radial excited state of the $X(3915)$ according to the mass  gap $M_{X(4500)}-M_{X(3915)}=588\,\rm{MeV}$.

The diquarks $q^{T}_j C\Gamma q^{\prime}_k$ have  five  structures  in Dirac spinor space, where $C\Gamma=C\gamma_5$, $C$, $C\gamma_\mu \gamma_5$,  $C\gamma_\mu $ and $C\sigma_{\mu\nu}$ for the scalar, pseudoscalar, vector, axialvector  and  tensor diquarks, respectively, the $j$ and $k$ are color indexes.
The attractive interactions of one-gluon exchange  favor  formation of
the diquarks in  color antitriplet $\varepsilon^{ijk} q^{T}_j C\Gamma q^{\prime}_k$ not in color sextet $d^{ajk} q^{T}_j C\Gamma q^{\prime}_k$ \cite{One-gluon}, where $a=1-6$, the structure constants $d^{ajk}=d^{akj}$,
  the favored configurations are the scalar ($C\gamma_5$) and axialvector ($C\gamma_\mu$) diquark states \cite{WangDiquark,WangLDiquark},
  the   heavy scalar and axialvector  diquark states have almost  degenerate masses from the QCD sum rules \cite{WangDiquark}.
We construct the  diquark-antidiquark type currents,
\begin{eqnarray}
&&C\gamma_5 \otimes \gamma_5C\, , \nonumber\\
&&C\gamma_\mu \otimes \gamma^\mu C\, ,
\end{eqnarray}
to study the lowest tetraquark states \cite{WangScalarT},  and observe that  the  $C\gamma_5\otimes \gamma_5 C$ type  and $C\gamma_\mu \otimes  \gamma^\mu C$  type hidden charm  tetraquark states have almost  degenerate masses \cite{WangMPLA,WangTetraquarkCTP}. In this article, we choose   the $C\gamma_\mu \otimes  \gamma^\mu C$ type current to study the $X(3915)$ and $X(4500)$ together.

In calculations, we observe that the lowest tetraquark masses are much larger than $5.0\,\rm{GeV}$,  if the $C\otimes C$ type interpolating currents are chosen.
The $C\gamma_\mu \gamma_5$ type diquark states are not as stable as the $C\gamma_\mu$ type and $C \gamma_5$ type diquark states,  the $C\gamma_\mu\gamma_5 \otimes \gamma_5 \gamma^\mu C$ type tetraquark states are expected to have much larger masses than that of the $C\gamma_5\otimes \gamma_5 C$ type  and $C\gamma_\mu \otimes  \gamma^\mu C$ type tetraquark states. So in this article, we choose the $C\gamma_\mu\gamma_5 \otimes \gamma_5 \gamma^\mu C$ type current to study the $X(4700)$.

In this article, we assign the  $X(3915)$ and $X(4500)$ to be the ground state and the first radial excited state of the $C\gamma_\mu \otimes \gamma^\mu C$ type $cs\bar{c}\bar{s}$ tetraquark states respectively, assign the $X(4700)$ to be the ground state of the $C\gamma_\mu\gamma_5 \otimes \gamma_5\gamma^\mu C$ type $cs\bar{c}\bar{s}$ tetraquark state, and study their  masses and pole residues with the QCD sum rules in details. In Ref.\cite{Chen1606}, Chen et al interpret the $X(4500)$  and $X(4700)$   as the D-wave  diquark-antidiquark type $cs\bar{c}\bar{s}$   tetraquark states with $J^P =0^+$,  the $X(4140)$  and $X(4274)$   as the S-wave  diquark-antidiquark type $cs\bar{c}\bar{s}$   tetraquark states with $J^P =1^+$ based on the QCD sum rules.

The article is arranged as follows:  we derive the QCD sum rules for the masses and pole residues of  the $X(3915)$, $X(4500)$ and $X(4700)$ in section 2; in section 3, we present the numerical results and discussions; section 4 is reserved for our conclusion.

\section{QCD sum rules for  the  $X(3915)$, $X(4500)$ and $X(4700)$ }
In the following, we write down  the two-point correlation functions $\Pi(p)$ and $\Pi_5(p)$ in the QCD sum rules,
\begin{eqnarray}
\Pi(p)&=&i\int d^4x e^{ip \cdot x} \langle0|T\left\{J(x)J^{\dagger}(0)\right\}|0\rangle \, , \\
\Pi_5(p)&=&i\int d^4x e^{ip \cdot x} \langle0|T\left\{J_5(x)J_5^{\dagger}(0)\right\}|0\rangle \, ,
\end{eqnarray}
where
\begin{eqnarray}
 J(x)&=&\varepsilon^{ijk}\varepsilon^{imn}s^T_j(x)C\gamma_\mu c_k(x)\, \bar{s}_m(x)\gamma^\mu C \bar{c}^T_n(x) \, ,  \nonumber \\
  J_5(x)&=&\varepsilon^{ijk}\varepsilon^{imn}s^T_j(x)C\gamma_\mu\gamma_5 c_k(x) \,\bar{s}_m(x)\gamma_5\gamma^\mu C \bar{c}^T_n(x) \, .
\end{eqnarray}
 We  choose the   currents $J(x)$ and $J_{5}(x)$  to interpolate the $X(3915)$, $X(4500)$ and $X(4700)$, respectively.

At the  phenomenological side, we  insert  a complete set of intermediate hadronic states with
the same quantum numbers as the current operators $J(x)$ and $J_5(x)$ into the
correlation functions $\Pi(p)$ and $\Pi_{5}(p)$ to obtain the hadronic representation
\cite{SVZ79,PRT85}. After isolating the ground state
and the first radial excited state contributions from the pole terms in the $\Pi(p)$, which are supposed to be the tetraquark states   $X(3915)$  and $X(4500)$ respectively, and isolating the ground state  contribution from the pole term in the $\Pi_5(p)$, which is supposed to be the tetraquark state  $X(4700)$,  we get the following results,
\begin{eqnarray}
\Pi(p)&=&\frac{\lambda_{X(3915)}^2}{M_{X(3915)}^2-p^2}+\frac{\lambda_{X(4500)}^2}{M_{X(4500)}^2-p^2}  +\cdots \, \, ,\\
\Pi_5(p)&=&\frac{\lambda_{X(4700)}^2}{M_{X(4700)}^2-p^2}  +\cdots \, \, ,
\end{eqnarray}
where the pole residues or coupling constants  $\lambda_{X(3915/4500/4700)}$ are defined by
\begin{eqnarray}
 \langle 0|J(0)|X(3915/4500)(p)\rangle &=&\lambda_{X(3915/4500)}  \, , \nonumber\\
 \langle 0|J_5(0)|X(4700)(p)\rangle &=&\lambda_{X(4700)}  \, .
\end{eqnarray}
There maybe also exist non-vanishing  coupling constants $\lambda^{\prime}_{X(3915/4500/4700)}$,
\begin{eqnarray}
 \langle 0|J_5(0)|X(3915/4500)(p)\rangle &=&\lambda^{\prime}_{X(3915/4500)}  \, , \nonumber\\
 \langle 0|J(0)|X(4700)(p)\rangle &=&\lambda^{\prime}_{X(4700)}  \, ,
\end{eqnarray}
we can take into account those contributions. In calculations, we observe that the existence  of the non-vanishing  coupling constants $\lambda^{\prime}_{X(3915/4500/4700)}$ leads to bad QCD sum rules. It is better to neglect them.

The tetraquark operators $J(x)$ and $J_5(x)$ contain a hidden strange component.  If we contract  the  quark pair $s,\,\bar{s}$ in the currents $J(x)$ and $J_5(x)$, and  substitute it by  the quark condensate $\langle \bar{s}s\rangle$, we obtain
 \begin{eqnarray}
 J(x)& \to& J^\prime(x)= \frac{2}{3}\langle \bar{s}s\rangle \, \bar{c}(x)c(x) \, ,  \nonumber \\
  J_5(x)& \to& J^\prime_5(x)= -\frac{2}{3}\langle \bar{s}s\rangle \, \bar{c}(x)c(x) \, .
 \end{eqnarray}
 The scalar currents $J^\prime(x)$ and $J^\prime_5(x)$ couple potentially to the scalar charmonium $\chi_{c0}(3414)$,
 \begin{eqnarray}
 \langle 0|J^\prime(0)|\chi_{c0}(p)\rangle&=&-\lambda^\prime_{\chi_{c0}}\, , \nonumber\\
 \langle 0|J_5^\prime(0)|\chi_{c0}(p)\rangle&=&\lambda^\prime_{\chi_{c0}}=-\frac{2}{3}\langle \bar{s}s\rangle f_{\chi_{c0}}M_{\chi_{c0}}\approx 0.9\times 10^{-2}\,\rm{GeV}^5\, ,
 \end{eqnarray}
 where  the decay constant $f_{\chi_{c0}}=359\,\rm{MeV}$ from the QCD sum rules \cite{Novikov-PRT1978}. The coupling constants have the relation  $\lambda^\prime_{\chi_{c0}} \ll \lambda_{X(3915/4500/4700)}$, moreover, the   $s$ and $\bar{s}$ in the currents $J(x)$ and $J_5(x)$  are valent quarks, while the   $s$ and $\bar{s}$ in the currents $J^\prime(x)$ and $J^\prime_5(x)$  are not valent quarks, they are just normalization factors. So the contaminations from the $\chi_{c0}(3414)$ are very small.

The diquark-antidiquark type currents  couple potentially    to  tetraquark states, the currents can be re-arranged both in the color and Dirac-spinor  spaces, and changed  to   special superpositions of   color  singlet-singlet type currents,
\begin{eqnarray}
J(x)&=&\bar{c}(x)c(x)\,\bar{s}(x)s(x)+\bar{c}(x)i\gamma_5c(x)\,\bar{s}(x)i\gamma_5s(x)+\frac{1}{2}\bar{c}(x)\gamma_{\alpha} c(x)\,\bar{s}(x)\gamma^{\alpha}s(x)\nonumber\\
&&-\frac{1}{2}\bar{c}(x)\gamma_{\alpha}\gamma_5 c(x)\,\bar{s}(x)\gamma^{\alpha}\gamma_5s(x)
 +\bar{c}(x)s(x)\,\bar{s}(x)c(x)+\bar{c}(x)i\gamma_5s(x)\,\bar{s}(x)i\gamma_5c(x)\nonumber\\
 &&+\frac{1}{2}\bar{c}(x)\gamma_{\alpha} s(x)\,\bar{s}(x)\gamma^{\alpha}c(x)-\frac{1}{2}\bar{c}(x)\gamma_{\alpha}\gamma_5 s(x)\,\bar{s}(x)\gamma^{\alpha}\gamma_5c(x) \, ,
\end{eqnarray}
\begin{eqnarray}
J_5(x)&=&-\bar{c}(x) i\gamma_5 c(x)\,\bar{s}(x)i\gamma_5s(x)-\bar{c}(x)c(x)\,\bar{s}(x)s(x)-\frac{1}{2}\bar{c}(x)\gamma_{\alpha} \gamma_5c(x)\,\bar{s}(x)\gamma^{\alpha}\gamma_5s(x)\nonumber\\
&&+\frac{1}{2}\bar{c}(x)\gamma_{\alpha}c(x)\,\bar{s}(x)\gamma^{\alpha}s(x)
 +\bar{c}(x)s(x)\,\bar{s}(x)c(x)+\bar{c}(x)i\gamma_5s(x)\,\bar{s}(x)i\gamma_5c(x)\nonumber\\
 &&-\frac{1}{2}\bar{c}(x)\gamma_{\alpha} s(x)\,\bar{s}(x)\gamma^{\alpha}c(x)+\frac{1}{2}\bar{c}(x)\gamma_{\alpha}\gamma_5 s(x)\,\bar{s}(x)\gamma^{\alpha}\gamma_5c(x) \, .
\end{eqnarray}
  The color  singlet-singlet type currents couple potentially to the meson-meson pairs or molecular states. The
diquark-antidiquark type tetraquark state can be taken as a special superposition of a series of  meson-meson pairs, and embodies  the net effects.
The component  $\bar{c}(x) c(x)\,\bar{s}(x) s(x)$ couples potentially to the meson pair $\chi_{c0}(3414)\,f_0(980)$, not the scalar charmonium $\chi_{c0}(3414)$ alone,  the main component of the $f_0(980)$ is $\bar{s}s$ from the QCD sum rules \cite{Wang-f0-980}. The contaminations from the $\chi_{c0}(3414)$ can be neglected safely.

 In the following,  we briefly outline  the operator product expansion for the correlation functions $\Pi(p)$ and $\Pi_5(p)$ in perturbative QCD.  We contract the $s$ and $c$ quark fields in the correlation functions
$\Pi(p)$ and $\Pi_5(p)$ with Wick theorem, and obtain the results:
\begin{eqnarray}
 \Pi(p)&=&i\varepsilon^{ijk}\varepsilon^{imn}\varepsilon^{i^{\prime}j^{\prime}k^{\prime}}\varepsilon^{i^{\prime}m^{\prime}n^{\prime}}\int d^4x e^{ip \cdot x}   \nonumber\\
&&{\rm Tr}\left[ \gamma_{\mu}C^{kk^{\prime}}(x)\gamma_{\alpha} CS^{jj^{\prime}T}(x)C\right] {\rm Tr}\left[ \gamma^{\alpha} C^{n^{\prime}n}(-x)\gamma^{\mu} C S^{m^{\prime}mT}(-x)C\right]   \, ,  \\
\Pi_5(p)&=&i\varepsilon^{ijk}\varepsilon^{imn}\varepsilon^{i^{\prime}j^{\prime}k^{\prime}}\varepsilon^{i^{\prime}m^{\prime}n^{\prime}}\int d^4x e^{ip \cdot x}   \nonumber\\
&&{\rm Tr}\left[ \gamma_{\mu}\gamma_5 C^{kk^{\prime}}(x)\gamma_5\gamma_{\alpha} CS^{jj^{\prime}T}(x)C\right] {\rm Tr}\left[ \gamma^{\alpha} \gamma_5C^{n^{\prime}n}(-x)\gamma_5\gamma^{\mu} C S^{m^{\prime}mT}(-x)C\right]   \, ,
\end{eqnarray}
where the   $S_{ij}(x)$   and $C_{ij}(x)$ are the full $s$ and $c$ quark propagators, respectively,
 \begin{eqnarray}
S_{ij}(x)&=& \frac{i\delta_{ij}\!\not\!{x}}{ 2\pi^2x^4}
-\frac{\delta_{ij}m_s}{4\pi^2x^2}-\frac{\delta_{ij}\langle
\bar{s}s\rangle}{12} +\frac{i\delta_{ij}\!\not\!{x}m_s
\langle\bar{s}s\rangle}{48}-\frac{\delta_{ij}x^2\langle \bar{s}g_s\sigma Gs\rangle}{192}+\frac{i\delta_{ij}x^2\!\not\!{x} m_s\langle \bar{s}g_s\sigma
 Gs\rangle }{1152}\nonumber\\
&& -\frac{ig_s G^{a}_{\alpha\beta}t^a_{ij}(\!\not\!{x}
\sigma^{\alpha\beta}+\sigma^{\alpha\beta} \!\not\!{x})}{32\pi^2x^2} -\frac{i\delta_{ij}x^2\!\not\!{x}g_s^2\langle \bar{s} s\rangle^2}{7776} -\frac{\delta_{ij}x^4\langle \bar{s}s \rangle\langle g_s^2 GG\rangle}{27648}-\frac{1}{8}\langle\bar{s}_j\sigma^{\mu\nu}s_i \rangle \sigma_{\mu\nu} \nonumber\\
&&   -\frac{1}{4}\langle\bar{s}_j\gamma^{\mu}s_i\rangle \gamma_{\mu }+\cdots \, ,
\end{eqnarray}
\begin{eqnarray}
C_{ij}(x)&=&\frac{i}{(2\pi)^4}\int d^4k e^{-ik \cdot x} \left\{
\frac{\delta_{ij}}{\!\not\!{k}-m_c}
-\frac{g_sG^n_{\alpha\beta}t^n_{ij}}{4}\frac{\sigma^{\alpha\beta}(\!\not\!{k}+m_c)+(\!\not\!{k}+m_c)
\sigma^{\alpha\beta}}{(k^2-m_c^2)^2}\right.\nonumber\\
&&\left. +\frac{g_s D_\alpha G^n_{\beta\lambda}t^n_{ij}(f^{\lambda\beta\alpha}+f^{\lambda\alpha\beta}) }{3(k^2-m_c^2)^4}-\frac{g_s^2 (t^at^b)_{ij} G^a_{\alpha\beta}G^b_{\mu\nu}(f^{\alpha\beta\mu\nu}+f^{\alpha\mu\beta\nu}+f^{\alpha\mu\nu\beta}) }{4(k^2-m_c^2)^5}+\cdots\right\} \, ,\nonumber\\
f^{\lambda\alpha\beta}&=&(\!\not\!{k}+m_c)\gamma^\lambda(\!\not\!{k}+m_c)\gamma^\alpha(\!\not\!{k}+m_c)\gamma^\beta(\!\not\!{k}+m_c)\, ,\nonumber\\
f^{\alpha\beta\mu\nu}&=&(\!\not\!{k}+m_c)\gamma^\alpha(\!\not\!{k}+m_c)\gamma^\beta(\!\not\!{k}+m_c)\gamma^\mu(\!\not\!{k}+m_c)\gamma^\nu(\!\not\!{k}+m_c)\, ,
\end{eqnarray}
and  $t^n=\frac{\lambda^n}{2}$, the $\lambda^n$ is the Gell-Mann matrix,  $D_\alpha=\partial_\alpha-ig_sG^n_\alpha t^n$ \cite{PRT85}. Then we compute  the integrals both in the coordinate space and in the momentum space to obtain the correlation functions $\Pi(p)$ and $\Pi_5(p)$ therefore the QCD spectral densities through dispersion relation.

In this article, we carry out the operator product expansion to the vacuum condensates up to dimension ($D$) 10 and
take the assumption of vacuum saturation for the  higher dimension vacuum condensates.
The condensates $\langle \frac{\alpha_s}{\pi}GG\rangle$, $\langle \bar{s}s\rangle\langle \frac{\alpha_s}{\pi}GG\rangle$,
$\langle \bar{s}s\rangle^2\langle \frac{\alpha_s}{\pi}GG\rangle$, $\langle \bar{s} g_s \sigma Gs\rangle^2$ and $g_s^2\langle \bar{s}s\rangle^2$ are the vacuum expectations of the operators of the order $\mathcal{O}(\alpha_s)$.   We take the truncations $D\leq 10$ and $k\leq 1$ in a consistent way,
the operators of the orders $\mathcal{O}( \alpha_s^{k})$ with $k> 1$ are  discarded.

 Finally  we can take the quark-hadron duality below the continuum thresholds  $s_X^0$ and perform Borel transform  with respect to
the variable $P^2=-p^2$ to obtain  the QCD sum rules:
\begin{eqnarray}
\lambda^2_{X(3915)}\, \exp\left(-\frac{M^2_{X(3915)}}{T^2}\right)+\lambda^2_{X(4500)}\, \exp\left(-\frac{M^2_{X(4500)}}{T^2}\right)&=& \int_{4m_c^2}^{s^0_{X(4500)}} ds\, \rho(s) \, \exp\left(-\frac{s}{T^2}\right) \, , \nonumber\\
\end{eqnarray}
\begin{eqnarray}
\lambda^2_{X(4700)}\, \exp\left(-\frac{M^2_{X(4700)}}{T^2}\right)&=& \int_{4m_c^2}^{s^0_{X(4700)}} ds\, \rho_5(s) \, \exp\left(-\frac{s}{T^2}\right) \, ,
\end{eqnarray}
where
\begin{eqnarray}
\rho_5(s)&=&\rho(s)\mid_{m_c \to -m_c}\, ,
\end{eqnarray}
the explicit expression of the QCD spectral density $\rho(s)$ is given in the appendix.

 We differentiate   Eq.(19) with respect to  $\frac{1}{T^2}$, then eliminate the
 pole residue $\lambda_{X(4700)}$, and  obtain the QCD sum rule for
 the mass of the  tetraquark state $X(4700)$,
 \begin{eqnarray}
 M^2_{X(4700)}= \frac{\int_{4m_c^2}^{s^0_{X(4700)}} ds\frac{d}{d \left(-1/T^2\right)}\rho_5(s)\exp\left(-\frac{s}{T^2}\right)}{\int_{4m_c^2}^{s^0_{X(4700)}} ds \rho_5(s)\exp\left(-\frac{s}{T^2}\right)}\, .
\end{eqnarray}
We take the predicted mass $M_{X(4700)}$ as input parameter, and obtain the pole residue  $\lambda_{X(4700)}$ from Eq.(19).

Now we study the masses and pole residues of the $X(3915)$ and $X(4500)$.
In Ref.\cite{Baxi-G}, M. S. Maior de Sousa and R. Rodrigues da Silva introduce a new approach  to calculate the masses and decay constants of the ground state and the first radial excited state of the conventional $\rho$, $\psi$ and $\Upsilon$ mesons with the QCD sum rules.
We introduce the notations $\tau=\frac{1}{T^2}$, $D^n=\left( -\frac{d}{d\tau}\right)^n$, and use the subscripts $1$ and $2$ to denote the ground state $X(3915)$ and the first radial excited state $X(4500)$, respectively, then write the QCD sum rule in Eq.(18) in the following form,
\begin{eqnarray}
\lambda_1^2\exp\left(-\tau M_1^2 \right)+\lambda_2^2\exp\left(-\tau M_2^2 \right)&=&\Pi_{QCD}(\tau) \, ,
\end{eqnarray}
where the subscript $QCD$  denotes the QCD side of the Borel transformed  correlation function.
We differentiate  the QCD sum rule with respect to $\tau$ to obtain
\begin{eqnarray}
\lambda_1^2M_1^2\exp\left(-\tau M_1^2 \right)+\lambda_2^2M_2^2\exp\left(-\tau M_2^2 \right)&=&D\Pi_{QCD}(\tau) \, .
\end{eqnarray}
Then we have two equations, it is easy to solve them to obtain the QCD sum rules,
\begin{eqnarray}
\lambda_i^2\exp\left(-\tau M_i^2 \right)&=&\frac{\left(D-M_j^2\right)\Pi_{QCD}(\tau)}{M_i^2-M_j^2} \, ,
\end{eqnarray}
where $i \neq j$.
Again we differentiate  above  QCD sum rules with respect to $\tau$ to obtain
\begin{eqnarray}
M_i^2&=&\frac{\left(D^2-M_j^2D\right)\Pi_{QCD}(\tau)}{\left(D-M_j^2\right)\Pi_{QCD}(\tau)} \, , \nonumber\\
M_i^4&=&\frac{\left(D^3-M_j^2D^2\right)\Pi_{QCD}(\tau)}{\left(D-M_j^2\right)\Pi_{QCD}(\tau)}\, .
\end{eqnarray}
 The squared masses $M_i^2$ satisfy the following equation,
\begin{eqnarray}
M_i^4-b M_i^2+c&=&0\, ,
\end{eqnarray}
where
\begin{eqnarray}
b&=&\frac{D^3\otimes D^0-D^2\otimes D}{D^2\otimes D^0-D\otimes D}\, , \nonumber\\
c&=&\frac{D^3\otimes D-D^2\otimes D^2}{D^2\otimes D^0-D\otimes D}\, , \nonumber\\
D^j \otimes D^k&=&D^j\Pi_{QCD}(\tau) \,  D^k\Pi_{QCD}(\tau)\, ,
\end{eqnarray}
$i=1,2$, $j,k=0,1,2,3$.
We solve the equation in Eq.(26) and obtain the solutions
\begin{eqnarray}
M_1^2=\frac{b-\sqrt{b^2-4c} }{2} \, , \\
M_2^2=\frac{b+\sqrt{b^2-4c} }{2} \, .
\end{eqnarray}
The squared masses $M_1^2$  and $M_2^2$ from the QCD sum rules in Eqs.(28-29) are functions of the Borel parameter $T^2$, continuum threshold parameter $s^0_{X}$ and energy scale $\mu$.

In Ref.\cite{Baxi-G}, M. S. Maior de Sousa and R. Rodrigues da Silva extract the masses and decay constants of the conventional mesons $\rho({\rm 1S,2S})$, $\psi({\rm 1S,2S})$, $\Upsilon({\rm 1S,2S})$ from the QCD spectral densities at the special  energy scales $\mu=1\,\rm{GeV}$, $m_c(m_c)$ and $m_b(m_b)$, respectively,  and observe that the theoretical values of the ground state masses are smaller than the experimental values. The new approach has a remarkable shortcoming.

In Ref.\cite{Wang4430}, we apply the new  approach  to study the hidden charm tetraquark states $Z_c(3900)$ and $Z(4430)$, and use the energy scale formula,
\begin{eqnarray}
\mu&=&\sqrt{M^2_{X/Y/Z}-(2{\mathbb{M}}_c)^2}\, ,
\end{eqnarray}
to determine the energy scales of the QCD spectral densities so as to overcome the shortcoming, and reproduce the experimental values of the masses $M_{Z_c(3900)}$ and $M_{Z(4430)}$ satisfactorily, where the $X/Y/Z$ denote the tetraquark states and the  ${\mathbb{M}}_c$ is the effective $c$-quark mass \cite{WangTetraquarkCTP,WangTetraquark2874}.  We  take the masses $M_{Z_c(3900)}$ and $M_{Z(4430)}$ from the BES collaboration and LHCb collaboration respectively as  input parameters to determine the optimal energy scales $\mu=\sqrt{M^2_{Z_c(3900)}-(2{\mathbb{M}}_c)^2}$, $\sqrt{M^2_{Z(4430)}-(2{\mathbb{M}}_c)^2}$  of the QCD spectral density firstly, then we search for the suitable Borel parameter and continuum threshold parameter, and obtain predicted masses $M_1$  and $M_2$ from the QCD sum rules, which happen to coincide with the experimental values $M_{Z_c(3900)}$ and $M_{Z(4430)}$, respectively. On the other hand, we  vary the energy scales $\mu$ of the QCD spectral density, and search for the suitable Borel parameter and continuum threshold parameter to extract the masses $M_1$  and $M_2$ at each energy scale. In calculations, we observe that the predicted masses  $M_1$  and $M_2$ vary with the energy scales $\mu$, the optimal energy scales $\mu$ satisfy the energy scale formula in Eq.(30) \cite{WangTetraquarkCTP,WangTetraquark2874,WangHuangTao-NPA,WangTetraquark1601}. The two routines lead to the same result, we can choose either of them.

In this article, we  choose the first routine, take the masses $M_{X(3915)}$, $M_{X(4500)}$ and $M_{X(4700)}$ from the Particle Data Group and LHCb collaboration respectively as  input parameters, use the energy scale formula in Eq.(30) to determine the energy scales of the QCD spectral densities and extract the masses $M_{1}$, $M_{2}$ and $M_{X(4700)}$ from Eq.(28),  Eq.(29) and Eq.(21), respectively, and examine whether or not they coincide with the experimental values $M_{X(3915)}=3918.4\,\rm{ MeV}$, $M_{X(4500)}=4506 \mbox{ MeV}$ and $M_{X(4700)}=4704 \mbox{ MeV}$ respectively, in other words, whether or not the predicted masses satisfy the energy scale formula.

\section{Numerical results and discussions}
The basic input parameters at the QCD side  are shown explicitly in Table 1.
The quark condensates, mixed quark condensates and $\overline{MS}$ masses  evolve according to the   renormalization group equation, we take into account
the energy-scale dependence,
\begin{eqnarray}
 \langle\bar{s}s \rangle(\mu)&=&\langle\bar{s}s \rangle(Q)\left[\frac{\alpha_{s}(Q)}{\alpha_{s}(\mu)}\right]^{\frac{4}{9}}\, , \nonumber\\
 \langle\bar{s}g_s \sigma Gs \rangle(\mu)&=&\langle\bar{s}g_s \sigma Gs \rangle(Q)\left[\frac{\alpha_{s}(Q)}{\alpha_{s}(\mu)}\right]^{\frac{2}{27}}\, , \nonumber\\
m_c(\mu)&=&m_c(m_c)\left[\frac{\alpha_{s}(\mu)}{\alpha_{s}(m_c)}\right]^{\frac{12}{25}} \, ,\nonumber\\
m_s(\mu)&=&m_s({\rm 2GeV} )\left[\frac{\alpha_{s}(\mu)}{\alpha_{s}({\rm 2GeV})}\right]^{\frac{4}{9}} \, ,\nonumber\\
\alpha_s(\mu)&=&\frac{1}{b_0t}\left[1-\frac{b_1}{b_0^2}\frac{\log t}{t} +\frac{b_1^2(\log^2{t}-\log{t}-1)+b_0b_2}{b_0^4t^2}\right]\, ,
\end{eqnarray}
  where $t=\log \frac{\mu^2}{\Lambda^2}$, $b_0=\frac{33-2n_f}{12\pi}$, $b_1=\frac{153-19n_f}{24\pi^2}$, $b_2=\frac{2857-\frac{5033}{9}n_f+\frac{325}{27}n_f^2}{128\pi^3}$,  $\Lambda=213\,\rm{MeV}$, $296\,\rm{MeV}$  and  $339\,\rm{MeV}$ for the flavors  $n_f=5$, $4$ and $3$, respectively  \cite{PDG}.

\begin{table}
\begin{center}
\begin{tabular}{|c|c|c|c|}\hline\hline
    Parameters                                          & Values\\   \hline
   $\langle\bar{q}q \rangle({\rm 1GeV})$                & $-(0.24\pm 0.01\, \rm{GeV})^3$ \,\, \cite{SVZ79,PRT85,ColangeloReview}         \\  \hline
   $\langle\bar{s}s \rangle({\rm 1GeV})$                & $(0.8\pm0.1)\langle\bar{q}q \rangle({\rm 1GeV})$ \,\, \cite{SVZ79,PRT85,ColangeloReview}     \\ \hline
$\langle\bar{s}g_s\sigma G s \rangle({\rm 1GeV})$       & $m_0^2\langle \bar{s}s \rangle({\rm 1GeV})$  \,\,  \cite{SVZ79,PRT85,ColangeloReview}        \\  \hline
$m_0^2({\rm 1GeV})$                                     & $(0.8 \pm 0.1)\,\rm{GeV}^2$      \,\,  \cite{SVZ79,PRT85,ColangeloReview}    \\   \hline
 $\langle \frac{\alpha_s GG}{\pi}\rangle$               & $(0.33\,\rm{GeV})^4$          \,\,  \cite{SVZ79,PRT85,ColangeloReview} \\   \hline
   $m_{s}({\rm 2GeV})$                                  & $(0.095\pm0.005)\,\rm{GeV}$  \,\, \cite{PDG}      \\   \hline
   $m_{c}(m_c)$                                         & $(1.275\pm0.025)\,\rm{GeV}$ \,\, \cite{PDG}      \\    \hline\hline
\end{tabular}
\end{center}
\caption{ The  basic input parameters in the QCD sum rules. }
\end{table}

In Refs.\cite{WangHuangTao-3900,WangTetraquarkCTP,WangTetraquark2874,WangHuangTao-NPA}, we  study the  hidden charm (bottom) tetraquark states  systematically  with the QCD sum rules by calculating the  vacuum condensates up to dimension 10  in
the operator product expansion in a consistent way,  and  explore the energy scale dependence of the hidden charm (bottom) tetraquark states  in details for the first time, and suggest a  formula
\begin{eqnarray}
\mu&=&\sqrt{M^2_{X/Y/Z}-(2{\mathbb{M}}_Q)^2} \, ,
 \end{eqnarray}
  to determine  the energy scales of the  QCD spectral densities. In Refs.\cite{WangHuangTao-3900,WangTetraquarkCTP,WangTetraquark2874}, we obtain the  effective mass  for  the diquark-antidiquark type tetraquark states   ${\mathbb{M}}_c=1.8\,\rm{GeV}$. Later, we re-checked the numerical calculations and found that there exists  a small error involving the mixed condensates.  We correct the small error, and obtain the optimal value  ${\mathbb{M}}_c=1.82\,\rm{GeV}$ \cite{WangTetraquark1601}. The Borel windows are modified slightly and the numerical results are also improved slightly. In this article, we choose the updated value ${\mathbb{M}}_c=1.82\,\rm{GeV}$.

 If we assign the $X(3915)$ and $X(4500)$ to be the ground state and the
  first radial excited state of the $C\gamma_\mu\otimes\gamma^\mu C$ type tetraquark states, respectively, the optimal energy scales are $\mu=1.45\,\rm{GeV}$ and $2.65\,\rm{GeV}$ for the QCD spectral density in the QCD sum rules for  the $X(3915)$ and $X(4500)$, respectively, the shortcoming of the new approach introduced  in Ref.\cite{Baxi-G} is overcome.
    At the energy scale $\mu=1.45\,\rm{GeV}$ and $2.65\,\rm{GeV}$, we can obtain the physical values $M_{X(3915)}$ and $M_{X(4500)}$ respectively, the associate values $M_{X(4500)}$ and $M_{X(3915)}$ from the coupled Eqs.(28-29) are not necessary the physical values, and are discarded. On the other hand, if we assign the $X(4700)$ to be the ground state $C\gamma_\mu\gamma_5\otimes\gamma_5\gamma^\mu C$ type tetraquark state, the optimal energy scale is $\mu=3.00\,\rm{GeV}$.

In the conventional QCD sum rules \cite{SVZ79,PRT85}, there are two criteria (pole dominance at the phenomenological side and convergence of the operator product
expansion at the QCD side) for choosing  the Borel parameters $T^2$ and continuum threshold parameters $s^0_X$.  Now we  search for  the Borel parameters $T^2$ and continuum threshold parameters $s_X^0$  to satisfy the two  criteria. The resulting Borel parameters and continuum threshold parameters are
\begin{eqnarray}
  X(3915/4500) &:& T^2 = (2.2-2.6) \mbox{ GeV}^2 \, ,\, s^0_{X(4500)} = (4.9\pm0.1 \mbox{ GeV})^2 \, , \\
   X(4700) &:& T^2 = (3.8-4.2) \mbox{ GeV}^2 \, ,\, s^0_{X(4700)} = (5.3\pm0.1 \mbox{ GeV})^2 \, .
\end{eqnarray}
The  contributions of the pole terms are
\begin{eqnarray}
  X(3915)+X(4500) &:& {\rm pole}  = (74-91) \% \, \,\, {\rm at}\, \,\, \mu = 1.45 \mbox{ GeV} \, , \\
    X(3915)+X(4500) &:& {\rm pole}  = (82-95) \% \, \,\, {\rm at}\, \,\, \mu = 2.65 \mbox{ GeV} \, , \\
   X(4700) &:& {\rm pole}  = (43-62) \% \, \,\, {\rm at}\, \,\, \mu = 3.00 \mbox{ GeV}   \, ,
\end{eqnarray}
the pole dominance at the phenomenological side is satisfied.
The contributions come from the vacuum condensates of dimension 10 $D_{10}$ are
\begin{eqnarray}
  X(3915)+X(4500) &:& D_{10}  = (1-3) \% \, \,\, {\rm at}\, \,\, \mu = 1.45 \mbox{ GeV} \, , \\
    X(3915)+X(4500) &:& D_{10}  = (1-2) \% \, \,\, {\rm at}\, \,\, \mu = 2.65 \mbox{ GeV} \, , \\
   X(4700) &:& D_{10}  \ll 1 \% \, \,\, {\rm at}\, \,\, \mu = 3.00 \mbox{ GeV}   \, ,
\end{eqnarray}
the operator product expansion at the QCD side is well convergent. So it is reliable  to extract the masses and pole residues from the QCD sum rules.

 Now we take into account the uncertainties of all the input parameters, and obtain the values of the masses and pole residues of the $X(3915)$, $X(4500)$ and $X(4700)$,
\begin{eqnarray}
M_{X(3915)}&=&3.92^{+0.19}_{-0.18}\,\rm{GeV} \, ,  \,\,\, {\rm Experimental\,\, value} \,\,\,\,3918.4\pm 1.9\,\rm{ MeV} \, \cite{PDG}\,   , \nonumber\\
M_{X(4500)}&=&4.83^{+1.32}_{-0.22}\,\rm{GeV} \, ,  \nonumber\\
\lambda_{X(3915)}&=&3.90^{+1.73}_{-1.12}\times 10^{-2}\,\rm{GeV}^5 \, , \nonumber\\
\lambda_{X(4500)}&=&1.01^{+5.05}_{-0.22}\times 10^{-1}\,\rm{GeV}^5 \,   ,
\end{eqnarray}
at the energy scale $\mu=1.45\,\rm{GeV}$,
\begin{eqnarray}
M_{X(3915)}&=&3.45^{+0.11}_{-0.10}\,\rm{GeV} \,   , \nonumber\\
M_{X(4500)}&=&4.50^{+0.08}_{-0.09}\,\rm{GeV} \, ,  \,\,\, {\rm Experimental\,\, value} \,\,\,\,4506 \pm 11 ^{+12}_{-15}\, \rm {MeV}\, \cite{LHCb2016}\,   , \nonumber\\
\lambda_{X(3915)}&=&2.64^{+0.47}_{-0.38}\times 10^{-2}\,\rm{GeV}^5 \, , \nonumber\\
\lambda_{X(4500)}&=&1.21^{+0.18}_{-0.14}\times 10^{-1}\,\rm{GeV}^5 \,   ,
\end{eqnarray}
at the energy scale $\mu=2.65\,\rm{GeV}$,
and
\begin{eqnarray}
M_{X(4700)}&=&4.70^{+0.08}_{-0.09}\,\rm{GeV} \, ,  \,\,\, {\rm Experimental\,\, value} \,\,\,\,4704 \pm 10 ^{+14}_{-24} \,\rm {MeV}\, \cite{LHCb2016}\,   , \nonumber\\
\lambda_{X(4700)}&=&1.47^{+0.24}_{-0.22}\times 10^{-1}\,\rm{GeV}^5 \,   ,
\end{eqnarray}
at the energy scale $\mu=3.00\,\rm{GeV}$. The energy scale formula in Eq.(30) is well satisfied.

\begin{figure}
\centering
\includegraphics[totalheight=6cm,width=7cm]{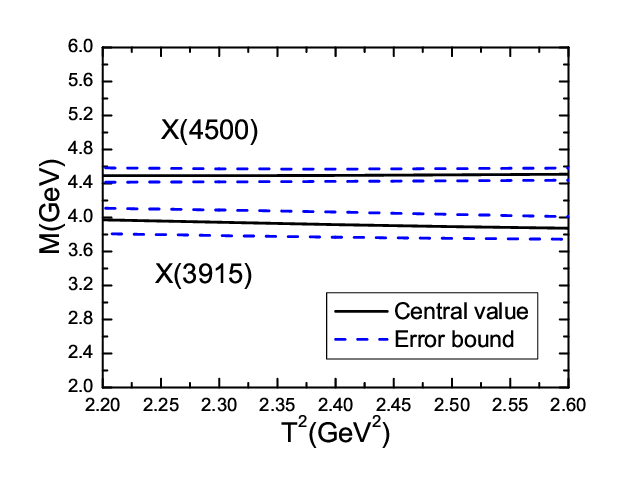}
\includegraphics[totalheight=6cm,width=7cm]{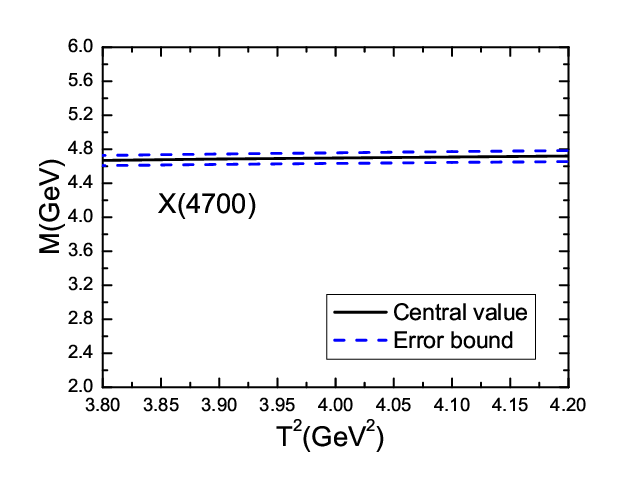}
  \caption{ The  masses  with variations of the  Borel parameters $T^2$. }
\end{figure}

Then we take the central values of the predicted masses and pole residues as the input parameters, and obtain the corresponding  pole contributions of the $X(3915)$ and $X(4500)$ respectively,
\begin{eqnarray}
{\rm pole}_{X(3915)}&=&(56-77)\% \, , \nonumber \\
{\rm pole}_{X(4500)}&=&(14-18)\% \, ,
\end{eqnarray}
at the energy scale $\mu=1.45\,\rm{GeV}$ and
\begin{eqnarray}
{\rm pole}_{X(3915)}&=&(44-65)\% \, , \nonumber \\
{\rm pole}_{X(4500)}&=&(30-38)\% \, ,
\end{eqnarray}
at the energy scale $\mu=2.65\,\rm{GeV}$.

It is more reliable to extract the  masses and pole residues from the QCD sum rules with larger pole contributions.
 The pole contribution of the $X(3915)$  at  $\mu=1.45\,\rm{GeV}$  is larger   than that at $\mu=2.65\,\rm{GeV}$, we prefer to extract the mass and pole residue of the $X(3915)$ at  $\mu=1.45\,\rm{GeV}$  and discard the ones at   $\mu=2.65\,\rm{GeV}$. On the other hand, the pole contribution of the $X(4500)$ at   $\mu=2.65\,\rm{GeV}$ is larger   than that at  $\mu=1.45\,\rm{GeV}$, we prefer to extract the mass and pole residue of the $X(4500)$ at  $\mu=2.65\,\rm{GeV}$ and discard the ones at   $\mu=1.45\,\rm{GeV}$.
 In this article, we take  the referred  values
  \begin{eqnarray}
M_{X(3915)}&=&3.92^{+0.19}_{-0.18}\,\rm{GeV} \, ,  \,\,\, {\rm Experimental\,\, value} \,\,\,\,3918.4\pm 1.9\,\rm{ MeV} \, \cite{PDG}\,   , \nonumber\\
M_{X(4500)}&=&4.50^{+0.08}_{-0.09}\,\rm{GeV} \, ,  \,\,\, {\rm Experimental\,\, value} \,\,\,\,4506 \pm 11 ^{+12}_{-15} \rm {MeV}\, \cite{LHCb2016}\,   , \nonumber\\
\lambda_{X(3915)}&=&3.90^{+1.73}_{-1.12}\times 10^{-2}\,\rm{GeV}^5 \, , \nonumber\\
\lambda_{X(4500)}&=&1.21^{+0.18}_{-0.14}\times 10^{-1}\,\rm{GeV}^5 \,   ,
\end{eqnarray}
  as the physical values. The predicted masses $M_{X(3915)}=3.92^{+0.19}_{-0.18}\,\rm{GeV}$,  $M_{X(4500)}=4.50^{+0.08}_{-0.09}\,\rm{GeV}$,  $M_{X(4700)}=4.70^{+0.08}_{-0.09}\,\rm{GeV}$  satisfy the energy scale formula in Eq.(30).

The predicted masses $M_{X(3915)}=3.92^{+0.19}_{-0.18}\,\rm{GeV}$,  $M_{X(4500)}=4.50^{+0.08}_{-0.09}\,\rm{GeV}$,  $M_{X(4700)}=4.70^{+0.08}_{-0.09}\,\rm{GeV}$,  which are shown explicitly in Fig.1,  are in excellent agreement with the experimental data, the present calculations support assigning the $X(3915)$ and $X(4500)$ to be the ground state and the
  first radial excited state of the $C\gamma_\mu\otimes\gamma^\mu C$ type $cs \bar{c}\bar{s}$ tetraquark states, and assigning the  $X(4700)$ to be the ground state $C\gamma_\mu\gamma_5\otimes\gamma_5\gamma^\mu C$ type $cs \bar{c}\bar{s}$ tetraquark state.

Now we study the finite width effects on the predicted tetraquark  masses. The   currents $J(x)$ and $J_{5}(x)$   couple potentially  with the scattering states  $J/\psi\phi$,
$D_s \bar{D}_s$, $D_s^* \bar{D}^*_s$, $\cdots$,  we take into account  the contributions of the  intermediate   meson-loops to the correlation functions $\Pi(p^2)$ and $\Pi_{5}(p^2)$,
\begin{eqnarray}
\Pi(p^2) &=&-\frac{\widehat{\lambda}_{X(3915)}^{2}}{ p^2-\widehat{M}_{X(3915)}^2-\Sigma^{X(3915)}_{J/\psi\phi}(p)+\cdots}-\frac{\widehat{\lambda}_{X(4500)}^{2}}{ p^2-\widehat{M}_{X(4500)}^2-\Sigma^{X(4500)}_{J/\psi\phi}(p)+\cdots}+\cdots \, , \nonumber\\
\Pi_{5}(p^2) &=&-\frac{\widehat{\lambda}_{X(4700)}^{2}}{ p^2-\widehat{M}_{X(4700)}^2-\Sigma^{X(4700)}_{J/\psi\phi}(p)+\cdots}+\cdots \, ,
\end{eqnarray}
where the $\widehat{\lambda}_{X(3915/4500/4700)}$ and $\widehat{M}_{X(3915/4500/4700)}$ are bare quantities to absorb the divergences in the self-energies $\Sigma^{X(3915/4500/4700)}_{J/\psi\phi}(p)$, $\cdots$.
All the renormalized self-energies  contribute  a finite imaginary part to modify the dispersion relation,
\begin{eqnarray}
\Pi(p^2) &=&-\frac{\lambda_{X(3915)}^{2}}{ p^2-M_{X(3915)}^2+i\sqrt{p^2}\Gamma_{X(3915)}(p^2)}-\frac{\lambda_{X(4500)}^{2}}{ p^2-M_{X(4500)}^2+i\sqrt{p^2}\Gamma_{X(4500)}(p^2)}+\cdots \, , \nonumber\\
\Pi_{5}(p^2) &=&-\frac{\lambda_{X(4700)}^{2}}{ p^2-M_{X(4700)}^2+i\sqrt{p^2}\Gamma_{X(4700)}(p^2)}+\cdots \, .
 \end{eqnarray}

We  take into account the finite width effects by the following simple replacements of the hadronic spectral densities,
\begin{eqnarray}
\delta \left(s-M^2_{X(3915/4500/4700)} \right) &\to& \frac{1}{\pi}\frac{\sqrt{s}\,\Gamma_{X(3915/4500/4700)}(s)}{\left(s-M_{X(3915/4500/4700)}^2\right)^2+s\,\Gamma_{X(3915/4500/4700)}^2(s)}\, ,
\end{eqnarray}
where
\begin{eqnarray}
\Gamma_{X(3915/4500/4700)}(s)&=&\Gamma_{X(3915/4500/4700)} \frac{M^2_{X(3915/4500/4700)} }{s } \, .
\end{eqnarray}
The experimental values of the widths are $\Gamma_{X(3915)}=20\pm5 \,\rm{MeV}$ \cite{PDG}, $\Gamma_{X(4500)}=92 \pm 21 ^{+21}_{-20} \,\rm{MeV}$, $\Gamma_{X(4700)}= 120 \pm 31 ^{+42}_{-33} \,\rm{MeV}$ \cite{LHCb2016}.

Then the phenomenological sides of  the QCD sum rules in Eqs.(18-19) undergo the following changes,
\begin{eqnarray}
B_T^2\Pi&=&\lambda^2_{X(3915)}\exp \left(-\frac{M^2_{X(3915)}}{T^2} \right)+\lambda^2_{X(4500)}\exp \left(-\frac{M^2_{X(4500)}}{T^2} \right) \nonumber\\
&\to& \frac{\lambda^2_{X(3915)}}{\pi}\int_{(M_{J/\psi}+M_{\omega})^2}^{s^0_{X(4500)}}ds\frac{\sqrt{s}\,\Gamma_{X(3915)}(s)}{(s-M_{X(3915)}^2)^2+s\Gamma_{X(3915)}^2(s)}\exp \left(-\frac{s}{T^2} \right)  \nonumber\\
&&+\frac{\lambda^2_{X(4500)}}{\pi}\int_{(M_{J/\psi}+M_{\phi})^2}^{s^0_{X(4500)}}ds\frac{\sqrt{s}\,\Gamma_{X(4500)}(s)}{(s-M_{X(4500)}^2)^2+s\Gamma_{X(4500)}^2(s)}\exp \left(-\frac{s}{T^2} \right)  \nonumber\\
&=&0.97\, \left\{ \lambda^2_{X(3915)}\exp \left(-\frac{M^2_{X(3915)}}{T^2} \right)+\lambda^2_{X(4500)}\exp \left(-\frac{M^2_{X(4500)}}{T^2} \right)\right\}\, , \\
B_T^2\Pi_5&=&\lambda^2_{X(4700)}\exp \left(-\frac{M^2_{X(4700)}}{T^2} \right) \nonumber\\
&\to& \frac{\lambda^2_{X(4700)}}{\pi}\int_{(M_{J/\psi}+M_{\phi})^2}^{s^0_{X(4700)}}ds\frac{\sqrt{s}\,\Gamma_{X(4700)}(s)}{(s-M_{X(4700)}^2)^2+s\Gamma_{X(4700)}^2(s)}\exp \left(-\frac{s}{T^2} \right)  \nonumber\\
&=&0.99\, \lambda^2_{X(4700)}\exp \left(-\frac{M^2_{X(4700)}}{T^2} \right)\, ,
\end{eqnarray}
and
\begin{eqnarray}
-\frac{1}{d(1/T^2)}B_T^2\Pi&=&M^2_{X(3915)}\lambda^2_{X(3915)}\exp \left(-\frac{M^2_{X(3915)}}{T^2} \right)+M^2_{X(4500)}\lambda^2_{X(4500)}\exp \left(-\frac{M^2_{X(4500)}}{T^2} \right) \nonumber\\
&\to& \frac{\lambda^2_{X(3915)}}{\pi}\int_{(M_{J/\psi}+M_{\omega})^2}^{s^0_{X(4500)}}ds\,s\,\frac{\sqrt{s}\,\Gamma_{X(3915)}(s)}{(s-M_{X(3915)}^2)^2+s\Gamma_{X(3915)}^2(s)}\exp \left(-\frac{s}{T^2} \right)  \nonumber\\
&&+\frac{\lambda^2_{X(4500)}}{\pi}\int_{(M_{J/\psi}+M_{\phi})^2}^{s^0_{X(4500)}}ds\,s\,\frac{\sqrt{s}\,\Gamma_{X(4500)}(s)}{(s-M_{X(4500)}^2)^2+s\Gamma_{X(4500)}^2(s)}\exp \left(-\frac{s}{T^2} \right)  \nonumber\\
&=&0.97\, \left\{ M^2_{X(3915)}\lambda^2_{X(3915)}\exp \left(-\frac{M^2_{X(3915)}}{T^2} \right)\right.\nonumber\\
&&\left.+M^2_{X(4500)}\lambda^2_{X(4500)}\exp \left(-\frac{M^2_{X(4500)}}{T^2} \right)\right\}\, , \\
-\frac{1}{d(1/T^2)} B_T^2\Pi_5&=&M^2_{X(4700)}\lambda^2_{X(4700)}\exp \left(-\frac{M^2_{X(4700)}}{T^2} \right) \nonumber\\
&\to& \frac{\lambda^2_{X(4700)}}{\pi}\int_{(M_{J/\psi}+M_{\phi})^2}^{s^0_{X(4700)}}ds\,s\,\frac{\sqrt{s}\,\Gamma_{X(4700)}(s)}{(s-M_{X(4700)}^2)^2+s\Gamma_{X(4700)}^2(s)}\exp \left(-\frac{s}{T^2} \right)  \nonumber\\
&=&0.99\, M^2_{X(4700)}\lambda^2_{X(4700)}\exp \left(-\frac{M^2_{X(4700)}}{T^2} \right)\, ,
\end{eqnarray}
where the $B_{T^2}$ denotes the Borel transformation. So we can absorb the numerical factors  $0.97$ and $0.99$ into the pole residues $\lambda_{X(3915/4500)}$ and $\lambda_{X(4700)}$ safely, the intermediate   meson-loops cannot  affect  the predicted masses $M_{X(3915/4500/4700)}$ significantly,
 the zero width approximation in  the phenomenological spectral densities   works.

\section{Conclusion}
In this article, we tentatively assign the  $X(3915)$ and $X(4500)$ to be ground state and the first radial excited state of the $C\gamma_\mu \otimes \gamma^\mu C$ type $cs\bar{c}\bar{s}$ tetraquark states, respectively, assign the $X(4700)$ to be the ground state $C\gamma_\mu\gamma_5 \otimes \gamma_5\gamma^\mu C$ type $cs\bar{c}\bar{s}$ tetraquark state, construct the corresponding interpolating currents,  and study their  masses and pole residues with the QCD sum rules  by calculating the contributions of the vacuum condensates up to dimension 10 in the operator product expansion. Moreover,  we use the  energy scale formula $\mu=\sqrt{M^2_{X/Y/Z}-(2{\mathbb{M}}_c)^2}$  to determine  the ideal energy scales of the QCD spectral densities.  The numerical results support assigning   the  $X(3915)$ and $X(4500)$ to be the ground state and the first radial excited state of the $C\gamma_\mu \otimes \gamma^\mu C$ type  $cs\bar{c}\bar{s}$ tetraquark states, respectively, and assigning  the $X(4700)$ to be the ground state of the $C\gamma_\mu\gamma_5 \otimes \gamma_5\gamma^\mu C$ type $cs\bar{c}\bar{s}$ tetraquark state.

\section*{Appendix}
The explicit expression of the QCD spectral density  $\rho(s)$,

\begin{eqnarray}
\rho(s)&=&\frac{1}{256\pi^6}\int_{y_i}^{y_f}dy \int_{z_i}^{1-y}dz \, yz\, (1-y-z)^3\left(s-\overline{m}_c^2\right)^2\left(7s^2-6s\overline{m}_c^2+\overline{m}_c^4 \right)  \nonumber\\
&&+\frac{1}{256\pi^6} \int_{y_i}^{y_f}dy \int_{z_i}^{1-y}dz \, yz \,(1-y-z)^2\left(s-\overline{m}_c^2\right)^3 \left(3s-\overline{m}_c^2\right)  \nonumber\\
&&+\frac{m_sm_c}{128\pi^6}\int_{y_i}^{y_f}dy \int_{z_i}^{1-y}dz \, (y+z)\, (1-y-z)^2 \left(s-\overline{m}_c^2\right)^2\left(5s-2\overline{m}_c^2 \right)  \nonumber\\
&&-\frac{m_c\langle \bar{s}s\rangle}{8\pi^4}\int_{y_i}^{y_f}dy \int_{z_i}^{1-y}dz \, (y+z)(1-y-z)\left(s-\overline{m}_c^2\right)\left(2s-\overline{m}_c^2\right)  \nonumber\\
&&+\frac{m_s\langle \bar{s}s\rangle}{8\pi^4}\int_{y_i}^{y_f}dy \int_{z_i}^{1-y}dz \, yz(1-y-z)\left(10s^2-12s\overline{m}_c^2+3\overline{m}_c^4 \right)   \nonumber\\
&&+\frac{m_s\langle \bar{s}s\rangle}{8\pi^4}\int_{y_i}^{y_f}dy \int_{z_i}^{1-y}dz \, yz\left(s-\overline{m}_c^2\right)\left(2s-\overline{m}_c^2\right)  \nonumber\\
&&-\frac{m_sm_c^2\langle \bar{s}s\rangle}{2\pi^4}\int_{y_i}^{y_f}dy \int_{z_i}^{1-y}dz  \left(s-\overline{m}_c^2\right)   \nonumber\\
&&-\frac{m_c^2}{192\pi^4} \langle\frac{\alpha_s GG}{\pi}\rangle\int_{y_i}^{y_f}dy \int_{z_i}^{1-y}dz \left( \frac{z}{y^2}+\frac{y}{z^2}\right)(1-y-z)^3 \nonumber\\
&&\left\{ 2s-\overline{m}_c^2+\frac{\overline{m}_c^4}{6}\delta\left(s-\overline{m}_c^2\right)\right\} \nonumber\\
&&-\frac{m_c^2}{384\pi^4}\langle\frac{\alpha_s GG}{\pi}\rangle\int_{y_i}^{y_f}dy \int_{z_i}^{1-y}dz \left(\frac{z}{y^2}+\frac{y}{z^2} \right) (1-y-z)^2 \left(3s-2\overline{m}_c^2\right) \nonumber\\
&&-\frac{1}{768\pi^4} \langle\frac{\alpha_s GG}{\pi}\rangle\int_{y_i}^{y_f}dy \int_{z_i}^{1-y}dz \left( y+z\right)(1-y-z)^2 \left( 10s^2-12s\overline{m}_c^2+3\overline{m}_c^4\right) \nonumber\\
&&+\frac{1}{384\pi^4} \langle\frac{\alpha_s GG}{\pi}\rangle\int_{y_i}^{y_f}dy \int_{z_i}^{1-y}dz \left( y+z\right)(1-y-z) \left( s-\overline{m}_c^2\right)\left( 2s-\overline{m}_c^2\right) \nonumber\\
&&+\frac{1}{384\pi^4} \langle\frac{\alpha_s GG}{\pi}\rangle\int_{y_i}^{y_f}dy \int_{z_i}^{1-y}dz \left( y+z\right)(1-y-z)^2 \left( 10s^2-12s\overline{m}_c^2+3\overline{m}_c^4\right) \nonumber
\end{eqnarray}

\begin{eqnarray}
&&+\frac{1}{3456\pi^4} \langle\frac{\alpha_s GG}{\pi}\rangle\int_{y_i}^{y_f}dy \int_{z_i}^{1-y}dz \, (1-y-z)^3 \left( 10s^2-12s\overline{m}_c^2+3\overline{m}_c^4\right)  \nonumber\\
&&+\frac{1}{576\pi^4} \langle\frac{\alpha_s GG}{\pi}\rangle\int_{y_i}^{y_f}dy \int_{z_i}^{1-y}dz  \, yz\,(1-y-z) \left( 10s^2-12s\overline{m}_c^2+3\overline{m}_c^4\right) \nonumber\\
&&+\frac{1}{576\pi^4} \langle\frac{\alpha_s GG}{\pi}\rangle\int_{y_i}^{y_f}dy \int_{z_i}^{1-y}dz \, (1-y-z)^2 \left( s-\overline{m}_c^2\right)
\left( 2s-\overline{m}_c^2\right) \nonumber\\
&&+\frac{1}{288\pi^4} \langle\frac{\alpha_s GG}{\pi}\rangle\int_{y_i}^{y_f}dy \int_{z_i}^{1-y}dz \, yz \left( s-\overline{m}_c^2\right)\left( 2s-\overline{m}_c^2\right) \nonumber\\
&&+\frac{m_c\langle \bar{s}g_s\sigma Gs\rangle}{32\pi^4}\int_{y_i}^{y_f}dy \int_{z_i}^{1-y}dz  \, (y+z) \left(3s-2\overline{m}_c^2 \right) \nonumber\\
&&-\frac{m_c\langle \bar{s}g_s\sigma Gs\rangle}{48\pi^4}\int_{y_i}^{y_f}dy \int_{z_i}^{1-y}dz  \,  (1-y-z) \left(3s-2\overline{m}_c^2 \right)     \nonumber\\
&&-\frac{m_s\langle \bar{s}g_s\sigma Gs\rangle}{8\pi^4}\int_{y_i}^{y_f}dy \int_{z_i}^{1-y}dz  \,  yz\left\{2s-\overline{m}_c^2+\frac{\overline{m}_c^2}{6} \delta\left(s-\overline{m}_c^2 \right) \right\}    \nonumber\\
&&-\frac{m_s\langle \bar{s}g_s\sigma Gs\rangle}{48\pi^4}\int_{y_i}^{y_f}dy    \,  y(1-y)\left(3s-2\widetilde{m}_c^2 \right)     \nonumber\\
&&+\frac{m_sm_c^2\langle \bar{s}g_s\sigma Gs\rangle}{8\pi^4}\int_{y_i}^{y_f}dy        \nonumber\\
&&-\frac{m_sm_c^2\langle \bar{s}g_s\sigma Gs\rangle}{48\pi^4}\int_{y_i}^{y_f}dy \int_{z_i}^{1-y}dz  \left(\frac{1}{y}+\frac{1}{z} \right) \nonumber\\
&&+\frac{m_c^2\langle\bar{s}s\rangle^2}{3\pi^2}\int_{y_i}^{y_f}dy   +\frac{g_s^2\langle\bar{s}s\rangle^2}{54\pi^4}\int_{y_i}^{y_f}dy \int_{z_i}^{1-y}dz\, yz \left\{2s-\overline{m}_c^2 +\frac{\overline{m}_c^4}{6}\delta\left(s-\overline{m}_c^2 \right)\right\}\nonumber\\
&&+\frac{g_s^2\langle\bar{s}s\rangle^2}{324\pi^4}\int_{y_i}^{y_f}dy \,y(1-y)\left(3s-2\widetilde{m}_c^2 \right)  \nonumber\\
&&-\frac{g_s^2\langle\bar{s}s\rangle^2}{648\pi^4}\int_{y_i}^{y_f}dy \int_{z_i}^{1-y}dz \, (1-y-z)\left\{ 3\left(\frac{z}{y}+\frac{y}{z} \right)\left(3s-2\overline{m}_c^2 \right)+\left(\frac{z}{y^2}+\frac{y}{z^2} \right)\right.\nonumber\\
&&\left.m_c^2\left[ 2+ \overline{m}_c^2\delta\left(s-\overline{m}_c^2 \right)\right]+(y+z)\left[12\left(2s-\overline{m}_c^2\right)+2\overline{m}_c^4\delta\left(s-\overline{m}_c^2\right) \right] \right\} \nonumber\\
&&-\frac{g_s^2\langle\bar{s}s\rangle^2}{1944\pi^4}\int_{y_i}^{y_f}dy \int_{z_i}^{1-y}dz \, (1-y-z)\left\{  15\left(\frac{z}{y}+\frac{y}{z} \right)\left(3s-2\overline{m}_c^2 \right)+7\left(\frac{z}{y^2}+\frac{y}{z^2} \right)\right. \nonumber\\
&&\left.m_c^2\left[ 2+\overline{m}_c^2\delta\left(s-\overline{m}_c^2\right)\right]+(y+z)\left[12\left(2s-\overline{m}_c^2\right) +2\overline{m}_c^4\delta\left(s-\overline{m}_c^2\right)\right] \right\} \nonumber\\
&&-\frac{m_sm_c \langle\bar{s}s\rangle^2}{12\pi^2}\int_{y_i}^{y_f}dy \left\{ 2+\widetilde{m}_c^2\delta(s-\widetilde{m}_c^2)\right\}\nonumber\\
&&+\frac{m_c^3\langle\bar{s}s\rangle}{144\pi^2  }\langle\frac{\alpha_sGG}{\pi}\rangle\int_{y_i}^{y_f}dy \int_{z_i}^{1-y}dz \left(\frac{y}{z^3}+\frac{z}{y^3}+\frac{1}{y^2}+\frac{1}{z^2}\right)(1-y-z)\nonumber\\
&&\left(1+\frac{ \overline{m}_c^2}{T^2}\right) \delta\left(s-\overline{m}_c^2\right)\nonumber\\
&&-\frac{m_c\langle\bar{s}s\rangle}{48\pi^2}\langle\frac{\alpha_sGG}{\pi}\rangle\int_{y_i}^{y_f}dy \int_{z_i}^{1-y}dz \left(\frac{y}{z^2}+\frac{z}{y^2}\right)(1-y-z)  \left\{2+\overline{m}_c^2\delta\left(s-\overline{m}_c^2\right) \right\}\nonumber
\end{eqnarray}

\begin{eqnarray}
&&+\frac{m_c\langle\bar{s}s\rangle}{48\pi^2}\langle\frac{\alpha_sGG}{\pi}\rangle\int_{y_i}^{y_f}dy \int_{z_i}^{1-y}dz\left\{2+ \overline{m}_c^2 \delta\left(s-\overline{m}_c^2\right) \right\} \nonumber\\
&&-\frac{m_c\langle\bar{s}s\rangle}{144\pi^2}\langle\frac{\alpha_sGG}{\pi}\rangle\int_{y_i}^{y_f}dy \int_{z_i}^{1-y}dz\left(\frac{1-y}{y}+\frac{1-z}{z}\right)
\left\{2+\overline{m}_c^2\delta\left(s-\overline{m}_c^2\right) \right\}\nonumber \\
&&-\frac{m_c\langle\bar{s}s\rangle}{288\pi^2}\langle\frac{\alpha_sGG}{\pi}\rangle\int_{y_i}^{y_f}dy \left\{2+ \widetilde{m}_c^2 \, \delta \left(s-\widetilde{m}_c^2\right) \right\}\nonumber\\
&&-\frac{m_c^2\langle\bar{s}s\rangle\langle\bar{s}g_s\sigma Gs\rangle}{6\pi^2}\int_0^1 dy \left(1+\frac{\widetilde{m}_c^2}{T^2} \right)\delta\left(s-\widetilde{m}_c^2\right)\nonumber \\
&&+\frac{ m_c^2\langle\bar{s}s\rangle\langle\bar{s}g_s\sigma Gs\rangle}{36\pi^2}\int_{0}^{1} dy \frac{1}{y(1-y)}\delta\left(s-\widetilde{m}_c^2\right) \nonumber\\
&&+\frac{m_c^2\langle\bar{s}g_s\sigma Gs\rangle^2}{48\pi^2T^6}\int_0^1 dy \, \widetilde{m}_c^4 \, \delta \left( s-\widetilde{m}_c^2\right)
\nonumber \\
&&-\frac{m_c^4\langle\bar{s}s\rangle^2}{54T^4}\langle\frac{\alpha_sGG}{\pi}\rangle\int_0^1 dy  \left\{ \frac{1}{y^3}+\frac{1}{(1-y)^3}\right\} \delta\left( s-\widetilde{m}_c^2\right)\nonumber\\
&&+\frac{m_c^2\langle\bar{s}s\rangle^2}{18T^2}\langle\frac{\alpha_sGG}{\pi}\rangle\int_0^1 dy  \left\{ \frac{1}{y^2}+\frac{1}{(1-y)^2}\right\} \delta\left( s-\widetilde{m}_c^2\right)\nonumber\\
&&+\frac{m_c^2\langle\bar{s}s\rangle^2}{54T^2}\langle\frac{\alpha_sGG}{\pi}\rangle\int_0^1 dy   \frac{1}{y(1-y)} \delta\left( s-\widetilde{m}_c^2\right)\nonumber \\
&&-\frac{m_c^2\langle\bar{s}g_s\sigma Gs\rangle^2}{144 \pi^2T^4} \int_0^1 dy   \frac{1}{y(1-y)}  \widetilde{m}_c^2 \, \delta\left( s-\widetilde{m}_c^2\right)\nonumber\\
&&+\frac{m_c^2\langle\bar{s}g_s\sigma Gs\rangle^2}{32 \pi^2T^2} \int_0^1 dy   \frac{1}{y(1-y)}   \delta\left( s-\widetilde{m}_c^2\right)\nonumber \\
&&+\frac{m_c^2\langle\bar{s} s\rangle^2}{54 T^6}\langle\frac{\alpha_sGG}{\pi}\rangle\int_0^1 dy \, \widetilde{m}_c^4 \, \delta \left( s-\widetilde{m}_c^2\right) \, ,
\end{eqnarray}
 where $y_{f}=\frac{1+\sqrt{1-4m_c^2/s}}{2}$,
$y_{i}=\frac{1-\sqrt{1-4m_c^2/s}}{2}$, $z_{i}=\frac{y
m_c^2}{y s -m_c^2}$, $\overline{m}_c^2=\frac{(y+z)m_c^2}{yz}$,
$ \widetilde{m}_c^2=\frac{m_c^2}{y(1-y)}$, $\int_{y_i}^{y_f}dy \to \int_{0}^{1}dy$, $\int_{z_i}^{1-y}dz \to \int_{0}^{1-y}dz$,
 when the $\delta$ functions $\delta\left(s-\overline{m}_c^2\right)$ and $\delta\left(s-\widetilde{m}_c^2\right)$ appear.

\section*{Acknowledgements}
This  work is supported by National Natural Science Foundation,
Grant Numbers 11375063,  and Natural Science Foundation of Hebei province, Grant Number A2014502017.

\end{document}